\documentclass[11pt]{article}

\usepackage[utf8]{inputenc}
\usepackage{authblk}
\usepackage[T1]{fontenc}
\usepackage{lmodern}
\usepackage[a4paper,margin=1in]{geometry}
\usepackage{amsmath,amssymb,amsthm}
\usepackage{booktabs,tabularx,array}
\usepackage{microtype}
\usepackage[authoryear,round]{natbib}
\usepackage[hidelinks]{hyperref}
\usepackage{enumitem}
\newcolumntype{Y}{>{\raggedright\arraybackslash}X}

\newcommand{\logit}{\operatorname{logit}}

\newcommand{\Prb}{\operatorname{Pr}}
\newcommand{\N}{\mathcal{N}}

\title{Differential Item Functioning: Methodological Advances in
Detection, Identification, and Interpretation}

\author[1]{Gabriel Wallin}
\author[2]{Qi Huang}

\affil[1]{School of Mathematical Sciences, Lancaster University}
\affil[2]{Department of Educational Studies, Purdue University}
\date{}

\begin{document}
\maketitle

\begin{abstract}
Differential item functioning (DIF) occurs when an item's response distribution differs across groups for respondents at the same level of the construct. Separating DIF from group differences in the construct itself requires identifying assumptions. Conventional analyses compare observed groups on a common scale established through items assumed to be invariant. We distinguish four settings according to whether the groups are observed or latent and whether the common scale is fixed by prespecified restrictions or estimated from the data. For observed groups, we review observed-score and item response theory procedures, multiple-covariate and intersectional models, and recursive partitioning. When anchors are unknown, we compare empirical anchor selection, partial and approximate invariance, robust scaling, regularisation and sparsity-based identification by the assumptions each makes about the pattern of DIF. For latent groups, we discuss mixture models and the joint estimation of class membership and DIF. Further topics include ordinal and multidimensional models, inference after data-dependent selection, effect sizes, consequences for composite scores, and the sources of DIF. We conclude with open problems, including anchor-free identification when DIF is widespread and inference when latent classes and invariant items are selected jointly.
\end{abstract}


\section{Introduction}

Comparisons of measurements across populations depend on whether an instrument represents the intended construct in a comparable way. At the item level, measurement invariance requires that respondents with the same standing on the intended construct have the same response distribution, irrespective of group membership. Differential item functioning (DIF) provides an item-level framework for investigating departures from this requirement. The relevant responses may be correct or incorrect answers, endorsements of symptoms or attitudes, reports of behaviour, or ordered ratings. The problem consequently arises across educational, psychological, clinical, health-related and other forms of multi-item measurement \citep{hollandthayer1988,swaminathan1990,millsap2011}.

DIF is not synonymous with a difference in observed response distributions. If groups have different distributions of the measured construct, their marginal probabilities may differ even when every item is invariant. The comparison of interest is therefore conditional on the intended construct. Nor is DIF synonymous with an unfair or biased item: a flagged item indicates a conditional discrepancy that calls for inspection of its content and measurement purpose. The distinction between differences in the latent construct distribution (termed \emph{impact} in much of the DIF literature) and differences in the measurement mechanism is thus fundamental.

Methodological work on DIF is extensive. Classical lines of research developed conditional odds-ratio and logistic-regression procedures, followed by tests of item parameter invariance within item response theory (IRT) \citep{hollandthayer1988,swaminathan1990,thissen1993}. Subsequent research addressed anchor-item selection, DIF associated with multiple observed covariates, latent subpopulations and increasingly complex measurement models \citep{bauer2023}. A systematic synthesis of simulation-based investigations by \citet{berrio2020}, for example, documents the breadth of DIF detection procedures and the conditions under which they have been studied. \citet{Zumbo2007} offers a complementary conceptual review, distinguishing three generations of DIF research and emphasising the need of explaining DIF through both item characteristics and broader contextual factors. Rather than attempting an exhaustive account of DIF methods, this paper examines the statistical foundations and selected methodological developments through a common identification problem: how can DIF be distinguished from population differences in the measured construct when the comparison groups, the invariant measurement structure, or both are uncertain?

Two distinct kinds of information enter many DIF analyses. The first is information about the populations across which item functioning is compared. Often the grouping variable is observed. Yet the groups associated with distinct response processes may be unobserved. For example, respondents may differ in response styles, interpretations of item wording or patterns of engagement in ways that were not directly recorded. Mixture IRT and related latent DIF approaches allow these groups to be estimated from response data \citep{rost1990,cohenbolt2005,cho2016}.

The second kind of information concerns how the common measurement scale is established. In multiple-group IRT, equality restrictions on suitable item parameters allow latent trait distributions to be compared. Items used for this purpose are commonly called anchors. When anchors are uncertain, DIF may be confounded with group differences in the latent trait, and an alternative source of identification is needed. Empirical anchor selection, regularisation, robust scaling, and partial or approximate invariance address this issue in different ways \citep{kopf2015a,belzak2020,chen2023,halpin2024}. In observed-score methods, where an explicit anchor set may not be specified, analogous concerns arise about whether the matching score adequately represents the construct without contamination from non-invariant items.

Crossing the two dimensions, observed or latent groups and prespecified or empirically established invariance, produces four methodological settings. This perspective is developed in \citet{wallin2024}, who also propose a regularised mixture IRT approach when both groups and anchors are unknown. The settings are not meant to be mutually exclusive categories. Rather, they distinguish the types of information an analysis relies on, and methods may combine assumptions from different settings.

Section~\ref{sec:framework} introduces a common statistical framework and examines the identification problem underlying
DIF analysis. Section~\ref{sec:observed} examines methods for observed groups, from conventional DIF procedures to approaches
that establish the invariant measurement structure empirically. Section~\ref{sec:latent} considers latent-group approaches and
the additional difficulties that arise when group membership and item non-invariance must be estimated simultaneously.
Section~\ref{sec:extensions} discusses extensions and practical considerations. The final section examines unresolved
methodological problems and directions for future research.

\section{Differential Item Functioning and the Identification of Measurement Non-Invariance}\label{sec:framework}

\subsection{Response functions, DIF effects and measurement scales}

Consider $N$ respondents and $J$ binary items. Write $Y_{ij}\in\{0,1\}$ for the response of respondent $i$ to item $j$, $\theta_i$ for the intended latent trait and $G_i\in\{0,1,\ldots,K\}$ for group membership. For the moment, assume that $G_i$ is observed. In a two-parameter logistic (2PL) IRT model, the reference-group response probability is
\begin{equation}\label{eq:2pl}
 p_{j0}(\theta)=\Prb(Y_{ij}=1\mid\theta_i=\theta,G_i=0)
 =\frac{\exp(a_j\theta+d_j)}{1+\exp(a_j\theta+d_j)},
\end{equation}
where $a_j>0$ is the item discrimination parameter and $d_j$ is the item intercept. The equivalent location parameter $b_j=-d_j/a_j$, often referred to as item difficulty in IRT, is the latent trait value at which the probability of a response coded as 1 equals $1/2$. The numerical values of the item parameters depend on the location and scale chosen for the latent trait.

Allowing item intercepts and discriminations to vary across groups gives
\begin{equation}\label{eq:general}
 \logit\,p_{jg}(\theta)
 =(a_j+\gamma_{jg})\theta+d_j+\delta_{jg},
 \qquad \gamma_{j0}=\delta_{j0}=0.
\end{equation}
Setting the reference-group deviations to zero defines the baseline parameterisation, but does not establish a common measurement scale across groups. The group-specific difference in the log odds of a response coded as 1 is
\begin{equation}\label{eq:difference}
 \Delta_{jg}(\theta)
 =\logit\,p_{jg}(\theta)-\logit\,p_{j0}(\theta)
 =\delta_{jg}+\gamma_{jg}\theta.
\end{equation}
When $\gamma_{jg}=0$, a non-zero $\delta_{jg}$ represents uniform DIF: the log odds ratio is constant across the latent trait. When $\gamma_{jg}\neq0$, the log odds ratio varies with $\theta$, indicating non-uniform DIF. Non-uniform DIF does not necessarily imply that the item
response curves cross within the substantively relevant trait range. Under this model, an item is invariant across all groups if
\begin{equation}\label{eq:null}
 H_{0j}:\quad\delta_{jg}=\gamma_{jg}=0
 \quad \text{for every }g=1,\ldots,K.
\end{equation}

The conditional nature of DIF is important. If the matching variable omits a relevant dimension of the intended construct, apparent group differences may arise even when a suitable multidimensional response model would be invariant \citep{ackermanma2024}. We return to this issue in Section~\ref{sec:extensions}.

We distinguish between invariant items and anchor items. An invariant item is one whose relevant response parameters are shared across groups under the specified model. An anchor item is an item used to establish or link the common measurement scale. An item selected as an anchor need not actually be invariant, and an invariant item need not be selected as an anchor. Approximate-invariance approaches may instead seek a sufficiently stable common measurement structure without requiring a distinct subset of exactly invariant items.

\subsection{Why impact and DIF are confounded}

The need for a common measurement scale can be illustrated using a uniform-DIF model. Suppose that
\begin{equation}\label{eq:locmodel}
 \logit\,p_{jg}(\theta)
 =a_j\theta+d_j+\delta_{jg},
\end{equation}
where $\delta_{j0}=0$ and the discrimination parameters are invariant across groups. Let $f_g(\theta)$ denote the latent trait density in group $g$, with mean $\mu_g$. Taking the reference-group metric as fixed, consider the group-specific transformation
\begin{equation}\label{eq:transform}
 \theta^*=\theta+c_g,\qquad
 \mu_g^*=\mu_g+c_g,\qquad
 \delta_{jg}^*=\delta_{jg}-a_jc_g,
\end{equation}
where $c_0=0$. Under this transformation,
\[
 a_j\theta^*+d_j+\delta_{jg}^*
 =a_j\theta+d_j+\delta_{jg}.
\]
If the latent trait distribution is free to shift accordingly, so that $f_g^*(t)=f_g(t-c_g)$, the marginal response distribution is unchanged. The observed responses therefore cannot distinguish
between parameterisations that differ only through this transformation. A group difference attributed to the latent trait mean under one parameterisation may instead be attributed, in part, to item-specific intercepts under another.

An anchor item provides a restriction that can resolve this ambiguity. If item $j$ is known to be invariant across groups, imposing $\delta_{jg}=0$ rules out a non-zero shift $c_g$, since $a_j>0$. Thus, with the reference-group metric fixed, a valid anchor can identify the relative latent trait location in this
simplified model. If the selected anchor is not actually invariant, however, the imposed restriction still determines a metric, but not necessarily the one intended. This distinction is central to the problem of anchor contamination.

When discrimination parameters are also allowed to vary across groups, the ambiguity extends to the scale of the latent trait. Writing the group-specific response function as
\[
 \logit\,p_{jg}(\theta)=a_{jg}\theta+d_{jg},
\]
consider the affine transformation
$\theta^*=s_g\theta+c_g$, where $s_g>0$. The same response function is obtained with
\begin{equation}\label{eq:affine}
 a_{jg}^*=\frac{a_{jg}}{s_g},\qquad
 d_{jg}^*=d_{jg}-\frac{a_{jg}c_g}{s_g}.
\end{equation}
Identifying a common metric therefore requires restrictions sufficient to determine both relative location and relative scale. The necessary restrictions depend on the response model, the assumptions about the latent trait distributions, and which item parameters are allowed to vary. Within-group identification conventions alone do not establish a common metric across groups.

Although individual DIF effects depend on the choice of metric, certain relative item comparisons do not. Under the Rasch model, with unit discrimination and group-specific intercepts
$d_{jg}=d_j+\delta_{jg}$, consider the contrast
\begin{equation}\label{eq:pair}
 (d_{jg}-d_{\ell g})
 -(d_{j0}-d_{\ell0})
 =\delta_{jg}-\delta_{\ell g}.
\end{equation}
A group-specific shift changes all item intercepts by the same amount, leaving their pairwise differences unchanged. Relative changes in item functioning can therefore be identified without assigning each individual DIF effect a unique location on the common scale. This is the basis of differential item-pair functioning approaches \citep{bechger2015}. A non-zero contrast establishes that the two items do not function invariantly relative to one another, but does not identify which item is responsible. Conversely, zero pairwise contrasts do not establish that both items are invariant on the intended common metric.

The following numerical example illustrates the consequences of imposing an incorrect anchor restriction. Consider five Rasch items with group-specific logit differences
\[
 (\delta_{1g},\ldots,\delta_{5g})
 =(0,0,0,0.6,0.8)
\]
on a common metric, with the reference-group latent mean fixed at zero. The first three items are invariant, whereas the fourth and fifth exhibit uniform DIF. Suppose that the fifth item is instead selected as an anchor. Setting $c_g=0.8$ in Equation~\eqref{eq:transform} gives the observationally equivalent parameterisation
\[
 (\delta_{1g}^*,\ldots,\delta_{5g}^*)
 =(-0.8,-0.8,-0.8,-0.2,0),
\]
while increasing the focal-group latent mean by $0.8$. The fifth item now appears invariant by construction, and the three originally invariant items exhibit apparent DIF in the opposite direction. The response distribution is identical under the two parameterisations; only the allocation of group differences between the latent trait and the item parameters has
changed.

The same example illustrates how a sparsity assumption can determine the common metric. Among the equivalent
parameterisations, consider choosing the shift that minimises the sum of absolute DIF effects,
\[
 \widehat c_g
 =\arg\min_{c_g}
 \sum_{j=1}^{5}|\delta_{jg}-c_g|.
\]
Any median of the five DIF effects minimises this objective. In the present example, the median is zero, so the original parameterisation is recovered. This result depends on the
configuration of DIF effects rather than on information in the response distribution alone. If the median DIF effect were non-zero, the same criterion would select a different origin.
Sparse identification thus replaces a prespecified anchor restriction with an assumption about the distribution of DIF
effects across items.

\subsection{The four methodological settings}

Table~\ref{tab:taxonomy} summarises the four methodological settings. The horizontal distinction concerns how the common measurement structure is established. Some methods rely on a prespecified matching variable, whereas others estimate or select the invariant structure from the data under additional assumptions. The vertical distinction concerns whether the groups
being compared are observed or inferred from the responses. Methods that identify subgroups from combinations of observed covariates, such as recursive partitioning, constitute another possibility (Section 3.2): group definitions are data-dependent, but the variables defining them are observed.

\begin{table}[htbp]
\centering\small
\caption{Four settings for DIF analysis according to information about groups and common measurement structure.}
\label{tab:taxonomy}

\begin{tabularx}{\textwidth}{p{0.18\textwidth}YY}
\toprule

& \textbf{Prespecified matching or linking structure}
& \textbf{Data-dependent identification of common measurement structure}\\

\midrule

\textbf{Observed groups}

& Conventional observed-score and multiple-group IRT procedures, including methods that use prespecified anchor constraints.

& Empirical anchor selection and purification; partial or approximate invariance; robust scaling; sparse or otherwise structured DIF estimation.\\[0.7em]

\textbf{Latent groups}

& Mixture IRT and related mixture measurement models with prespecified identifying constraints on item parameters or latent trait distributions.

& Joint estimation of latent group structure and item-specific differences under explicit identifying restrictions, such as sparsity or constraints on the distribution of DIF effects.\\

\bottomrule
\end{tabularx}
\end{table}

The four settings are not mutually exclusive categories of DIF methods. Observed-score procedures may use a total score without specifying an anchor set, but their validity still depends on the
adequacy of the matching variable. Bayesian approximate-invariance methods may assign prior distributions that concentrate item parameter differences near zero without requiring any item to be exactly invariant. Mixture IRT models may
incorporate observed covariates, and latent classes may differ in their trait distributions or response styles as well as their item parameters. The classification therefore concerns the role of
observed and estimated information in establishing group comparisons, rather than the particular estimation procedure used. In all four settings, the assumptions that establish a common measurement scale must be distinguished from those used to
detect differences in item functioning.

\section{DIF Analysis with Observed Groups}
\label{sec:observed}

\subsection{Prespecified group comparisons}

When group membership is defined in advance, DIF is assessed by comparing item responses across groups at comparable levels of the intended construct. The grouping variable may be, for example, gender, race, language, or an experimentally defined condition. Classical procedures condition on an observed matching variable, such as the total score or a rest score obtained by excluding the item under investigation, whereas latent variable models condition on a model-based latent trait.  

Mantel--Haenszel (MH), SIBTEST, and logistic regression are some of the most established observed-score procedures for DIF detection. The MH approach estimates a common conditional odds ratio across matching-score strata, which is particularly suited to detecting approximately uniform conditional differences, but does not explicitly model how the group contrast may vary with the matching variable \citep{hollandthayer1988}. SIBTEST estimates a weighted difference in expected item or subtest responses and can be applied to individual items or item bundles \citep{shealystout1993}. For an item bundle, the bundle score replaces the item score. The adequacy of the matching subtest and the regression correction remains central to interpretation. Logistic regression extends the MH approach by allowing the conditional group contrast to depend on the matching variable \citep{swaminathan1990}. By including a group main effect and a group-by-score interaction, the model provides tests of uniform and non-uniform DIF, respectively. Statistical significance should be considered alongside the magnitude of the estimated effects and the corresponding differences in predicted response probabilities.

MH, SIBTEST and logistic regression all depend on the adequacy of the matching variable; none resolves the underlying measurement problem automatically. An observed composite score is a proxy for the intended construct, and its suitability depends on the items from which it is formed. The matching score thus plays the role of the anchor. The total score is calculated from all the items, and the rest score from all items except the studied one. If several items exhibit DIF, conditioning on the composite score could produce misleading group comparisons. A rest score removes the studied item from the matching variable, but does not eliminate non-invariance among the remaining items. Moreover, conditioning on an observed score is not equivalent to conditioning on the latent trait itself. Under the Rasch model, the total score is sufficient for $\theta$, so when every item is DIF-free, the conditional response probabilities given the total score do not depend on the groups' trait distributions. Under e.g., the 2PL model, the total score is not sufficient and this property does not hold.  Purification procedures attempt to improve the matching variable by iteratively excluding items identified as potentially non-invariant. This is a form of anchor selection (Section 3.3), and it works only if the items left in the matching score are invariant.

Latent variable approaches define the matching variable through a measurement model. In multiple-group IRT, item parameters are compared across prespecified groups after the groups have been placed on a common latent metric. Likelihood-ratio tests compare constrained and less restricted model fits, Wald tests evaluate estimated parameter differences relative to their estimated covariance matrix, and score tests assess whether relaxing the null constraints would improve the local likelihood \citep{thissen1993,belzak2020}. In the 2PL item parameterisation, intercept differences represent uniform DIF, whereas discrimination differences indicate non-uniform DIF. A common latent
metric is typically established by fixing the mean and variance of the reference group while allowing them to vary in other groups, and constraining anchor-item parameters across groups \citep{belzak2020}. DIF detection and estimation remain relative to these identifying restrictions.

Residual-based IRT methods for DIF detection (RDIF) retain the conditioning on latent trait but change the diagnostic target. Rather than estimate group-specific item parameters, the RDIF fits an aggregate IRT model and compares group means of raw and squared item residuals: $\mathrm{RDIF}_{R}$ primarily targets uniform DIF, $\mathrm{RDIF}_{S}$ targets non-uniform DIF, and $\mathrm{RDIF}_{RS}$ provides a joint test \citep{lim2022}. GRDIF extends these residual comparisons to omnibus tests across multiple groups, followed by pairwise comparisons for flagged items \citep{lim2024}. Both use the aggregate calibration, which treats every item as invariant, as their reference structure; when purification is used, trait estimates are recomputed after provisionally flagged items are removed.

The multiple-indicator multiple-cause (MIMIC) model provides another
way to examine DIF when group membership or other relevant
covariates are observed. The model allows latent trait differences and uniform DIF to be represented separately \citep{woods2009mimic}. As in multiple-group IRT, items without direct covariate effects act as anchors. The conventional MIMIC specification assumes invariant discrimination parameters, and therefore only captures uniform DIF. Non-uniform DIF requires an extension that allows the relationship between the latent trait and the item response to differ across groups, for example through a group-by-trait interaction \citep{woodsgrimm2011}. Multiple-group IRT models represent DIF through group-specific item parameters, whereas MIMIC represents the corresponding differences as covariate effects. Explanatory item response modelling uses the same regression logic through interactions between an item indicator and a person covariate \citep{deboeckwilson2004}, providing a natural bridge to the multiple-covariate methods considered in Section~\ref{sec:observed_subgroups}.

\subsection{Multiple covariates, intersectional comparisons, and subgroup discovery}
\label{sec:observed_subgroups}

Observed-group DIF analyses do not need to be limited to comparisons based on a single grouping variable. Although conventional DIF analyses often examine one characteristic (e.g., gender, race/ethnicity, or language) at a time, individuals could simultaneously occupy multiple social positions \citep{crenshaw1989, bauer2014}. Therefore, single-axis comparisons may obscure DIF associated with particular combinations of characteristics. For instance, an item may show little DIF when gender and language background are examined separately, yet function differently for multilingual girls than would be expected from the separate gender and language effects. Recent literature has accordingly begun to formalise intersectional DIF as variation in item parameters associated with combinations of observed characteristics. This broader perspective motivates methods that model several observed covariates jointly or search the covariate space to discover subgroups with non-invariant item parameters.

In terms of Table 1, these methods are developed for observed groups. When the covariates and their interactions are prespecified, the groups are also prespecified. When subgroups are found by searching the covariate space, the group labels are data-dependent, although the variables defining them are observed. In both cases, item parameters can be compared across the identified groups as long as a common metric is established. 

Moderated nonlinear factor analysis (MNLFA) connects the multiple-group and MIMIC formulations while allowing more general patterns of observed heterogeneity \citep{bauerhussong2009,bauer2017}. By allowing several categorical or continuous covariates, including interactions, MNLFA moderates the latent trait mean and variance as well as item intercepts and discriminations. Covariate effects on the trait distribution represent impact, whereas effects on item parameters represent DIF conditional on the latent trait. For a dichotomous item, a simplified item response function can be written as
\begin{equation}
\operatorname{logit}\Pr(Y_{ij}=1\mid\theta_i,\mathbf{x}_i)
=\left(a_j+\mathbf{x}_i^{\top}\boldsymbol{\alpha}_j\right)\theta_i
+d_j+\mathbf{x}_i^{\top}\boldsymbol{\delta}_j,
\label{eq:mnlfa_dif}
\end{equation}
where $\mathbf{x}_i$ may contain multiple covariates, non-linear transformations, and interactions. Moderation of the latent mean or variance represents population differences in the construct. By contrast, $\boldsymbol{\delta}_j$ represents moderation of the item intercept and hence uniform DIF, whereas $\boldsymbol{\alpha}_j$ represents moderation of discrimination and hence non-uniform DIF. MNLFA can therefore estimate several sources of DIF without dividing the sample into separately calibrated groups, but the functional forms and interactions must be specified by the analyst \citep{bauer2023}.

A difference among cross-classified groups does not necessarily constitute intersectional DIF. For two binary characteristics, $A_i$ and $B_i$, intersectional uniform DIF is represented by their interaction in the item intercept after the additive effects of $A_i$ and $B_i$ have been accounted for. This distinction matters because an item difference observed for multilingual girls may reflect the separate effects of gender and language, rather than an additional effect associated with their combination. 

Although this fixed-effect representation is straightforward for a small number of prespecified covariates, it becomes difficult to specify and estimate as the number of covariates increases. Even if cross-classified strata can be specified, some may also contain few respondents. The multilevel random item effects model \citep{ren2026} offers a hierarchical alternative for this setting. Let $s$ index intersectional strata and let $\eta_{js}$ denote an item parameter for item $j$ in stratum $s$. Its central decomposition can be represented schematically as
\begin{equation}
\eta_{js}=\mathbf{X}_s^{\top}\boldsymbol{\beta}_j+u_{js},
\qquad
u_{js}\sim \mathcal{N}(0,\omega_j^2),
\label{eq:random_intersectional_dif}
\end{equation}
where $\mathbf{X}_s^{\top}\boldsymbol{\beta}_j$ represents additive demographic main effects and $u_{js}$ represents the remaining stratum deviation for item $j$. A non-zero $\omega_j^2$ indicates residual variation across identity combinations after the main effects have been accounted for. Regularisation is used to identify the item-specific variance components supported by the data. The principal benefit of this formulation is partial pooling: deviations for sparsely represented strata are shrunk more strongly toward the main effects prediction, which can stabilise estimation when many intersectional strata are present.

Tree-based methods address a different form of uncertainty by discovering rather than prespecifying subgroups. Score-based tests of parameter instability provide a way to assess invariance along an observed covariate, including a continuous or
ordinal one, without first dividing respondents into groups \citep{merklezeileis2013}. Tests of this kind supply the splitting criterion in model-based recursive partitioning. Rasch trees use model-based recursive partitioning to identify covariate-defined subgroups with different item parameters \citep{strobl2015rasch}. Item-focused trees instead fit partitions for individual items, allowing different covariates and cutpoints to account for non-invariance in different items \citep{tutzberger2016trees}. The latter approach has also been extended to ordinal responses under the partial credit model \citep{bollmann2018trees}. These methods should not be mistaken for mixture IRT: the subgroup boundaries are learned, but membership is defined by recorded characteristics rather than an unobserved class. Inference must account for the search over candidate splits.

The three methods for multiple covariates differ mainly in what must be specified. MNLFA tests covariate effects and interactions chosen in advance. The multilevel model uses partial pooling to estimate variation across intersectional groups after accounting for demographic main effects. Tree-based methods instead search for the covariates and cut-points that define groups with different item parameters.

\subsection{Establishing a common metric when anchor items are uncertain}
\label{sec:observed_metric}

The observed-group methods reviewed above all require a defensible basis to compare respondents at the same level of the intended construct. Observed-score procedures obtain this basis from a matching variable, whereas latent-variable approaches establish a common metric through identifying constraints, often equality restrictions on anchor-item parameters. The preceding discussion largely treated this basis as given; this subsection considers how a common metric can be established when invariant anchor items are not known in advance.

The location-shift equivalence established in Equation~\eqref{eq:transform} explains why incorrectly selected anchor items can distort DIF estimates. Under the uniform-DIF model in Equation~\eqref{eq:locmodel}, suppose that an item with $\delta_{jg}\neq0$ is incorrectly constrained to be invariant. If this item alone determines the relative latent trait location, the common metric can shift to satisfy the imposed restriction, redistributing the apparent DIF effects across the remaining items. In the Rasch model, where all discriminations equal one, the transformation is
\begin{equation}\label{eq:contam}
 \delta_{jg}^{*}=\delta_{jg}-c_g.
\end{equation}
An incorrectly selected anchor can therefore conceal DIF in some items and produce apparent DIF in others. With unequal discriminations, the induced change in the item intercepts depends on $a_j$, as shown in Equation~\eqref{eq:transform}. If several items are simultaneously constrained to be invariant, an incorrect anchor restriction cannot generally be accommodated by a single location shift without violating the remaining restrictions. The resulting model may then exhibit parameter bias or misfit. The choice of anchors thus affects both the estimated group differences and the item-level DIF conclusions.

When invariant items are not known in advance, empirical anchor-selection procedures attempt to identify a suitable reference set. Some procedures begin with a provisional anchor, remove items showing evidence of non-invariance, and repeat the analysis until the selected set stabilises; others rank candidate items using preliminary DIF statistics or construct an anchor by adding items sequentially. \citet{woods2009anchor} examine empirical rank-based anchor selection, while \citet{kopf2015a,kopf2015b} distinguish different anchor classes and selection strategies and investigate their statistical properties. An iterative selection procedure can be represented as
\begin{equation}\label{eq:purify}
 \mathcal A^{(t+1)}
 =\mathcal S\left(
 \{T_j(\mathcal A^{(t)}):j=1,\ldots,J\}
 \right),
\end{equation}
where $\mathcal A^{(t)}$ denotes the provisional anchor set at iteration $t$, $T_j(\mathcal A^{(t)})$ is the DIF statistic for item $j$ computed using that set, and $\mathcal S$ is the selection rule. Specific procedures differ in their initial anchor, the items included when each candidate is tested, and the rules used to update or terminate selection. These choices matter because the current reference set determines the metric on which the item statistics are calculated, and those statistics in turn determine the next reference set. Convergence alone does not establish that the resulting anchors are invariant.

Anchor length introduces a related trade-off. A short anchor containing suitable items may reduce exposure to contamination, but provide limited information for linking the latent scales. Including additional invariant
items can improve linking precision, whereas including non-invariant items may distort the common metric. One alternative approach is to provisionally use every item except the one under investigation. Constant-anchor methods retain a selected subset across item-level tests, while iterative-forward procedures construct the anchor by adding candidates according to a
selection criterion. Their statistical properties depend on the number of items and respondents, the prevalence and magnitude of DIF, and whether the item-specific effects are in the same or opposing directions \citep{kopf2015a,kopf2015b}.

Related identification problems arise in multiple-group factor analysis \citep{millsap2011}. Under partial measurement invariance, selected equality restrictions on item parameters are relaxed while sufficient constraints are kept to
establish a common latent metric \citep{byrne1989,millsap2011}. The relationship to DIF analysis is particularly close for categorical outcomes, where factor loadings correspond to discrimination parameters and thresholds are related to item locations. Partial invariance and empirical anchor selection thus address related measurement problems, although they need not employ the same constraints or selection procedures \citep{thissen2025}.

Identifying a subset of exactly invariant items is not the only way to establish a common measurement structure. Approximate-invariance approaches allow item parameters to differ across groups while placing restrictions on the magnitude or distribution of those differences. In a Bayesian formulation, for example, the group-specific DIF effects may be assigned priors of the form
\begin{equation}\label{eq:bayes}
 \delta_{jg}\sim\N(0,\omega^2),
\end{equation}
where a small positive value of $\omega^2$ expresses a prior preference for small departures from invariance. Unlike an exact anchor restriction, this specification does not require any particular item to have $\delta_{jg}=0$. The prior contributes information about the location and dispersion of the DIF effects, and the posterior could depend on the chosen prior variance. Sensitivity to this specification is therefore important, particularly when the responses provide limited information about the common metric \citep{muthen2012,vandeschoot2013}.

Alignment provides another approach to establishing a common metric without requiring exact invariance for a prespecified set of items. It estimates group locations and scales by seeking a parameterisation
that yields a relatively simple pattern of non-invariance, and has been developed particularly for comparisons involving many groups
\citep{muthenasparouhov2014}. Its identifying assumptions differ from those of exact anchor selection and Bayesian approximate invariance. These approaches should therefore not be treated as interchangeable solutions to anchor uncertainty. In particular, approximate invariance  must be assessed relative to the intended measurement comparison; it should not be treated as proof that every item functions identically.

Regularisation provides a third strategy by estimating DIF effects while shrinking selected parameters to zero. Let $\boldsymbol\psi$ collect the model parameters and let $\boldsymbol\delta_j=(\delta_{j1},\ldots,\delta_{jK})^\top$ contain the uniform DIF effects for item $j$. A general penalised estimator takes the form
\begin{equation}\label{eq:penalty}
 \widehat{\boldsymbol\psi}_{\lambda}
 =\underset{\boldsymbol\psi\in\Psi}{\arg\min}
 \left\{-\ell(\boldsymbol\psi)
 +\lambda\sum_{j=1}^J P(\boldsymbol\delta_j)\right\},
\end{equation}
where $\ell(\boldsymbol\psi)$ is the log-likelihood, $P$ is a penalty function and $\lambda$ controls the degree of regularisation. In a two-group model, the choice $P(\boldsymbol\delta_j)=|\delta_{j1}|$ gives an $\ell_1$ penalty, which can produce exactly zero estimates for individual DIF effects. With multiple groups, the group penalty $P(\boldsymbol\delta_j)=\|\boldsymbol\delta_j\|_2$ encourages all uniform DIF effects for an item to be selected or excluded together. An elementwise
penalty, $\sum_{g=1}^K|\delta_{jg}|$, instead allows individual group--item contrasts to be selected. If the purpose is to identify items exhibiting either uniform or non-uniform DIF, the penalty
must also account for the discrimination deviations $\gamma_{jg}$. An item with $\widehat{\boldsymbol\delta}_j=\boldsymbol 0$ is not necessarily invariant if its discrimination parameters differ across groups.

Penalised DIF estimation was introduced by \citet{magis2015}, who applied a lasso penalty to item-specific group effects so that DIF detection and anchor determination occur within a single fit. It has since been developed for models with multiple candidate covariates and
extended to different item response models.
\citet{tutz2015} consider a Rasch formulation with multiple potential sources of DIF, while \citet{belzak2020} use regularisation to select
anchor items and identify DIF in the 2PL model. Related procedures have also been developed for polytomous responses \citep{schauberger2020}. For multiple observed background variables, \citet{bauerbelzakcole2020} extend sparse DIF estimation to MNLFA, penalising covariate effects on item intercepts and loadings to select a common measurement structure without sequential item-by-item testing. This extends regularisation beyond comparisons defined by a single binary group variable. Rather than specifying an invariant subset before estimation, these methods impose restrictions on the collective pattern of item-specific deviations. Their success depends on the extent to which that pattern is consistent with the chosen penalty and the assumptions used to identify the common measurement scale.

Sparsity is particularly relevant when the invariant items are unknown, but it should not be confused with evidence that the estimated zero effects are exactly zero in the population. Nor does regularisation eliminate the need for identification assumptions. If DIF is widespread, the parameterisation favoured by the penalty may differ from the intended common metric. Alternative constraints, such as requiring the DIF effects to sum to zero, impose different assumptions about the overall pattern of non-invariance. A sum-to-zero restriction fixes an average deviation at zero, whereas an $\ell_1$ criterion favours a parameterisation in which the absolute deviations are collectively small. Neither restriction is justified by the response data alone.

The distinction between regularisation for parameter selection and an identifying condition based on sparsity can be seen from the location-shift equivalence class in Equation~\eqref{eq:transform}. Suppose that the discrimination parameters are invariant and positive, and that the
group-specific latent trait distributions are free to shift. Among
equivalent parameterisations, consider the location satisfying
\begin{equation}\label{eq:minl1}
 \widehat c_g\in
 \underset{c\in\mathbb R}{\arg\min}
 \sum_{j=1}^J|\delta_{jg}-a_jc|.
\end{equation}
Because
\[
 |\delta_{jg}-a_jc|
 =a_j\left|\frac{\delta_{jg}}{a_j}-c\right|,
\]
any weighted median of the ratios $\delta_{jg}/a_j$, with weights $a_j$, minimises the objective. The minimum need not be unique. In the Rasch model, where the discriminations equal one, the criterion reduces to the ordinary median of the DIF effects. The minimum-$\ell_1$ condition therefore provides a way to select a location within an otherwise equivalent set of parameterisations. Whether this location corresponds to
the intended common metric depends on the underlying configuration of DIF.
\citet{chen2023} develop an identification condition based on the minimum-$L_1$ norm and use it to construct estimates, confidence intervals and hypothesis tests without requiring a prespecified anchor set. The identifying condition is conceptually distinct from using a penalty to select zero coefficients during estimation.

Other methods avoid selecting a conventional anchor set by changing the quantities used to assess DIF  or the way the common metric is estimated. Differential item-pair functioning examines relative item locations, which remain identified under the location shifts discussed earlier \citep{bechger2015}. \citet{yuan2021} develop an approach
based on relative changes in item location differences, using QQ plots and a graphical test to examine item-level DIF. \citet{wang2022robust} define item-specific DIF effects through residuals from regressions relating item parameters across groups and use robust estimation
to reduce sensitivity to non-invariant items. \citet{halpin2024} formulates DIF as an outlier problem in cross-group IRT scaling and develops robust estimators of the linking parameters. These approaches differ in their estimands, estimation procedures and inferential properties. They also rely on different assumptions
about the shared relationships among items. Avoiding a prespecified anchor set does not remove the need for assumptions that establish a common latent metric.

When the invariant structure is estimated from the responses, a further difficulty concerns statistical inference. Penalised estimators are
generally biased by shrinkage, and the selected set
\[
 \widehat{\mathcal A}_{\lambda}
 =\{j:\widehat{\boldsymbol\delta}_j
 =\boldsymbol 0\}
\]
depends on the same observations used to estimate the item parameters. Confidence intervals and hypothesis tests calculated as though this set had been specified in advance do not typically have their nominal coverage or error rates. The uncertainty introduced by selecting the tuning
parameter may also need to be considered. Refitting the selected model without a penalty can reduce shrinkage bias, but does not by itself account for the selection process. Procedures for inference after data-dependent identification must therefore account for the conditions under which the measurement structure was estimated. The approach of \citet{chen2023} addresses this distinction by developing inference under an explicit identification condition, rather than treating the output
of an anchor-selection procedure as fixed.

When group membership is also unknown, item selection and latent classification become interdependent. This issue is examined in the following section.

\section{DIF Analysis with Latent Groups}
\label{sec:latent}

\subsection{Mixture measurement models and the meaning of latent DIF}

Observed grouping variables may not fully capture heterogeneity in item response processes. Respondents with the same recorded characteristics may differ in their response styles, interpretation of item wording, engagement or other processes that affect their responses. Mixture IRT models allow such heterogeneity to be represented through an unobserved categorical variable
$C_i\in\{0,\ldots,K\}$. Individual class membership is inferred from the responses rather than supplied before analysis, while the number of classes may be specified or selected through model comparison. The mixture Rasch model of \citet{rost1990} provides an early example of this approach. Subsequent work has applied mixture IRT to DIF analysis, including comparisons between observed and latent group membership
\citep{cohenbolt2005,cho2016}. A related approach in the latent class logistic regression framework of \citet{zumbo2015}, which allows DIF effecrs to vary across latent classes. 

In mixture IRT models, the presence of latent classes does not, by itself, establish DIF. Classes may differ in their latent trait distributions while sharing the same item response functions. It is therefore important to distinguish heterogeneity in the measured construct from class-specific differences in item functioning. For a binary-item model with uniform DIF, let
\begin{align}
 \Prb(C_i=g)
 &=\pi_g,\qquad
 \pi_g>0,\quad
 \sum_{g=0}^{K}\pi_g=1,
 \label{eq:mixturepi}\\
 \theta_i\mid C_i=g
 &\sim F_g,
 \label{eq:mixturetrait}\\
 p_{jg}(\theta)
 &=\operatorname{expit}
 (a_j\theta+d_j+\delta_{jg}),
 \qquad \delta_{j0}=0,
 \label{eq:mixtureitem}
\end{align}
where $F_g$ denotes the latent trait distribution in class $g$, with density $f_g$, and $\operatorname{expit}(x)=\exp(x)/\{1+\exp(x)\}$.
Assuming local independence conditional on $\theta_i$ and $C_i$, the marginal probability of a response pattern $\mathbf y_i$ is
\begin{equation}\label{eq:mixturelikelihood}
 \Prb(\mathbf Y_i=\mathbf y_i)
 =\sum_{g=0}^{K}\pi_g\int
  \prod_{j=1}^{J}
  p_{jg}(\theta)^{y_{ij}}
  \{1-p_{jg}(\theta)\}^{1-y_{ij}}
  f_g(\theta)\,d\theta.
\end{equation}
A common parametric specification assumes $F_g=\N(\mu_g,\sigma_g^2)$, with the reference-class location and scale fixed. The suitability of class-specific normal distributions and the number of mixture components are modelling questions rather than consequences of the mixture formulation itself.

Equation~\eqref{eq:mixturelikelihood} separates two sources of heterogeneity. The distributions $F_g$ describe differences in the latent trait across classes, whereas the parameters $\delta_{jg}$ describe class-specific departures from the common item response functions. When $K=0$, the model reduces to an ordinary IRT model. With multiple classes but
$\delta_{jg}=0$ for all $j$ and $g$, the measurement model is invariant and the classes differ only in their latent trait distributions. If some $\delta_{jg}\neq0$, respondents at the same latent trait level may have different response probabilities depending on class membership. Such differences constitute latent DIF when the item parameters are expressed on
an appropriately identified common metric. Allowing unrestricted class-specific item parameters without establishing that metric does not, by itself, permit meaningful comparisons of item functioning.

Mixture IRT is particularly relevant when the characteristics associated with differential functioning were not recorded or do not correspond closely to an observed grouping variable. \citet{cohenbolt2005}, for example, demonstrate that latent groups associated with differential functioning may correspond only weakly to the observed characteristic used in a conventional DIF analysis. \citet{deboeckchowilson2011} take a different approach, representing
latent DIF through a secondary dimension on which classes differ and using item covariates to explain which items are affected. The interpretation of these classes nevertheless requires information beyond the response patterns. Differences in item functioning may be consistent with distinct response
styles, interpretations of item content or other unobserved characteristics, but the fitted mixture does not establish which process produced them. Alternative explanations may also arise from misspecified latent trait distributions, unmodelled dimensions or violations of local independence.

When suitable invariant items are known in advance, they can be used to establish a common measurement scale across the latent classes. In the uniform-DIF model, an anchor set $\mathcal A$ imposes \begin{equation}\label{eq:latentanchor}
 \delta_{jg}=0
 \quad\text{for all }j\in\mathcal A
 \text{ and }g=1,\ldots,K.
\end{equation}
If non-uniform DIF is also permitted, the corresponding discrimination
deviations must be constrained for the anchor items. With the reference-class metric fixed, suitable cross-class restrictions can identify the relative locations and scales needed for item parameter comparisons. These
restrictions do not, however, guarantee identification of the entire mixture model. The number of classes, their separability and the information available for estimating class-specific parameters introduce additional identification and estimation difficulties. Class labels are also arbitrary unless they are given meaning through external information
or an explicit labelling convention.

\subsection{Learning both the groups and the invariant structure}

When neither group membership nor the invariant item structure is known, both must be inferred from the response data. These are not independent estimation problems. Posterior class membership depends on the class-specific latent trait distributions and item response functions, while estimates of item non-invariance depend on how respondents are assigned probabilistically to classes. Different assumptions about the invariant structure may therefore yield different latent classifications. Conversely, changes in the estimated class structure may alter which items are identified as exhibiting DIF. The model must distinguish differences in the latent trait distributions from class-specific differences in item functioning without relying on prespecified anchors or observed group labels.

One approach combines mixture IRT with penalised estimation. Under the uniform-DIF model introduced in Equation~\eqref{eq:mixtureitem}, consider the penalised marginal likelihood
\begin{equation}\label{eq:jointpenalty}
 \widehat{\boldsymbol\psi}_{\lambda}
 =\underset{\boldsymbol\psi\in\Psi}{\arg\min}
 \left\{-\sum_{i=1}^{N}
 \log\left[\sum_{g=0}^{K}\pi_g
 \int\prod_{j=1}^{J}
 p_{jg}(\theta)^{Y_{ij}}
 \{1-p_{jg}(\theta)\}^{1-Y_{ij}}
 f_g(\theta)\,d\theta\right]
 +\lambda\sum_{j=1}^{J}\sum_{g=1}^{K}|\delta_{jg}|
 \right\}.
\end{equation}
This formulation combines the estimation of latent classes with regularisation of item-specific DIF effects. \citet{wallin2024} develop an approach of this kind for the 2PL model, using an $L_1$ penalty and a tailored expectation--maximisation algorithm. Under the elementwise penalty in Equation~\eqref{eq:jointpenalty}, individual class-specific deviations may be estimated as zero. An item is selected as invariant across all classes only when every corresponding deviation is zero. Replacing the penalty with $\sum_{j=1}^{J}\|\boldsymbol\delta_j\|_2$
would instead encourage the selection of entire items as invariant or non-invariant. If class-specific discrimination parameters are also permitted, the penalty must be extended to the corresponding slope deviations. The choice of penalty thus determines both the structure of the estimated DIF effects and the level at which invariance is selected.

Related approaches differ in their measurement models and regularisation assumptions. \citet{robitzsch2022} investigate regularised mixture Rasch modelling using the smoothly clipped absolute deviation (SCAD) penalty
on class-specific item parameter differences. Unlike the convex $L_1$ penalty, SCAD is non-convex and is designed to reduce shrinkage of sufficiently large effects. Bayesian formulations offer another possibility, using prior distributions to represent assumptions about item invariance and latent class structure.  \citet{chenlumoustaki2022}, for example, place latent classes on both respondents and items to detect item leakage and
preknowledge; although framed as outlier detection, the model has the same structure as the setting in which both the affected groups and the non-invariant items are unknown. Shrinkage priors and hierarchical specifications can be used to favour parsimonious patterns of non-invariance, although the resulting estimates and posterior probabilities depend on the prior assumptions.

The joint identification problem extends beyond continuous-trait IRT. \citet{song2026cdm} develop an adaptive-LASSO approach for simultaneously identifying latent subgroups and DIF items within cognitive diagnostic assessment. The underlying attributes are discrete rather than continuous, but assumptions connecting common item functioning and population structure remain necessary. Extensions involving ordinal and asymmetric item response functions are considered below.

The identifying role of regularisation requires particular care when the groups are latent. Under the uniform-DIF model in Equation~\eqref{eq:mixtureitem}, consider a class-specific shift $\theta^*=\theta+c_g$, with $c_0=0$.
The transformation
\[
 \delta_{jg}^*=\delta_{jg}-a_jc_g,
 \qquad
 f_g^*(t)=f_g(t-c_g)
\]
leaves the class-specific marginal response distribution unchanged.
Consequently, the mixture distribution is also unchanged. If the DIF effects satisfy $\delta_{jg}=a_jc_g$ for all items in a class, they can be absorbed entirely into a shift of that class's  latent trait distribution. More generally, a penalty may allocate the component of the DIF pattern that can be represented as a latent location shift to the trait
distribution, while retaining the remaining item-specific deviations.
The resulting parameterisation is determined partly by the identifying
criterion. If the population DIF pattern does not satisfy the assumptions
underlying that criterion, the selected metric does not need to correspond to the intended measurement comparison.

A further complication arises because the latent classes have no predetermined labels. Permuting the class labels, together with their mixing proportions, latent trait distributions and item response parameters, leaves the mixture distribution unchanged. This label symmetry must be distinguished from the location and scale indeterminacies discussed earlier. The choice of reference class also 
requires attention when the penalty is imposed on deviations from its
item parameters. Relabelling the classes may change the reference
parameterisation and, consequently, the magnitude of the penalised
deviations. An elementwise penalty on reference-class contrasts is
therefore not necessarily invariant to the choice of reference class.
The identifying constraints and penalty must be considered together
when interpreting the estimated common measurement structure. Statistical separation of mixture components does not establish their substantive meaning. External information is needed to investigate what the classes represent.

\subsection{Estimation, class uncertainty and model assessment}

Estimation of mixture DIF models requires integration over both the unobserved group membership and the latent trait. For fixed model parameters, define the class-specific marginal likelihood of respondent
$i$ as
\begin{equation}\label{eq:lig}
 L_{ig}=\int\prod_{j=1}^{J}
 p_{jg}(\theta)^{Y_{ij}}
 \{1-p_{jg}(\theta)\}^{1-Y_{ij}}
 f_g(\theta)\,d\theta.
\end{equation}
The posterior probability that respondent $i$ belongs to class $g$ is
\begin{equation}\label{eq:tau}
 \tau_{ig}=\Prb(C_i=g\mid\mathbf Y_i)
 =\frac{\pi_gL_{ig}}
 {\sum_{h=0}^{K}\pi_hL_{ih}}.
\end{equation}
Conditional on class membership, the posterior density of the latent trait is
\[
 f(\theta\mid\mathbf Y_i,C_i=g)
 =
 \frac{
 f_g(\theta)
 \prod_{j=1}^{J}
 p_{jg}(\theta)^{Y_{ij}}
 \{1-p_{jg}(\theta)\}^{1-Y_{ij}}
 }{L_{ig}}.
\]
An EM algorithm uses these posterior quantities to calculate the expected complete-data log-likelihood in the E-step. The M-step updates the class proportions, latent trait distribution parameters, common item parameters and class-specific DIF effects. When the class proportions do not depend on covariates, the update is $\widehat\pi_g=N^{-1}\sum_i\tau_{ig}$. When DIF parameters are regularised, the M-step must also accommodate the penalty, potentially
through a nested optimisation procedure. \citet{wallin2024} develop a regularised EM approach for the joint estimation of latent groups and DIF structure.

The posterior probabilities in Equation~\eqref{eq:tau} quantify uncertainty
about individual class membership under the fitted model. Assigning each respondent to the class with the largest posterior probability and subsequently treating those assignments as observed groups discards this uncertainty. When the class probabilities are diffuse, such a procedure may distort estimates of class-specific characteristics and subsequent
analyses of item functioning. Class proportions and summaries of posterior membership probabilities provide a more appropriate description of the estimated population structure. These quantities should nevertheless
be interpreted conditionally on the fitted model: the classification does not establish that the classes correspond to substantively distinct populations or response mechanisms. External covariates may help characterise the classes, but their use should be distinguished from
including them in the class membership model.

Mixture estimation also presents numerical difficulties. The likelihood may have multiple local maxima, and different initial values can produce solutions with different class proportions, item parameters or selected
DIF effects. Model fitting should therefore use multiple initialisations and examine the stability of the maximised objective, boundary estimates, class membership probabilities and selected invariant items. When comparing
solutions, differences attributable only to permutations of the class labels must be distinguished from genuine differences in the estimated model. Small or poorly separated classes may provide limited information for estimating class-specific item parameters, even when the numerical
optimisation appears to have converged.

Selecting the number of classes introduces a further source of uncertainty. In regularised mixture models, the number of components and the penalty parameter both affect the estimated DIF structure.
Information criteria may help compare candidate specifications, but their use requires care when parameters are penalised or mixture components are weakly identified. In particular, the effective number of
parameters does not need to coincide with the number of non-zero estimates, and an information criterion favouring a more
complex model does not establish the substantive meaning of its additional
classes. The stability of the selected model across initialisations and plausible alternative specifications is therefore
important alongside numerical fit.

Likelihood-based inference can also be non-regular. When testing whether an
additional mixture component is needed, the null hypothesis might place a mixing proportion on the boundary of the parameter space, while parameters
associated with the additional component are not identified. Conventional
chi-square likelihood-ratio approximations are generally invalid in this setting. Tests of item parameter restrictions within a fixed, identifiable mixture may have different asymptotic properties, depending on the hypothesis and the remaining model constraints. Regularised selection introduces additional uncertainty because the estimated invariant structure depends on the responses and the selected tuning parameter. Bootstrap or other model-specific calibration procedures could be useful, but resampling must repeat the relevant class and item selection steps if the resulting
uncertainty estimates are intended to account for them. 

When assessing the model, one should examine whether class-specific item functioning is needed to represent the observed responses, rather than treating the presence of multiple latent classes as evidence of DIF. Where appropriate, a useful comparison involves three specifications: a conventional IRT model with invariant item response functions, a mixture
model with class-specific latent trait distributions but invariant item response functions, and a mixture model that additionally permits selected item parameters to differ across classes.

An important qualification concerns the interpretation of the second
specification. When the item response functions are invariant, integrating
over the latent classes gives
\[
\begin{aligned}
 \Prb(\mathbf Y_i=\mathbf y_i)
 &=
 \sum_{g=0}^{K}\pi_g
 \int
 \Prb(\mathbf Y_i=\mathbf y_i\mid\theta)
 f_g(\theta)\,d\theta\\
 &=
 \int
 \Prb(\mathbf Y_i=\mathbf y_i\mid\theta)
 \bar f(\theta)\,d\theta,
\end{aligned}
\]
where
\[
 \bar f(\theta)
 =\sum_{g=0}^{K}\pi_gf_g(\theta).
\]
Thus, a mixture with invariant item response functions is observationally equivalent to a single-class measurement model with the corresponding mixture distribution for the latent trait. An improvement over a model
that assumes a single normal trait distribution could therefore reflect a
more flexible representation of the latent distribution rather than distinct
measurement mechanisms. Allowing class-specific item parameters addresses
a different question: whether the response data support item-specific
heterogeneity beyond that represented by the assumed latent trait distribution.

\section{Extensions and Interpretations}
\label{sec:extensions}

The preceding sections describe how DIF can be defined and detected when groups are observed or latent and when the common metric is known or estimated. However, a valid interpretation requires more than DIF detection. The response model must be appropriate for the items and the intended construct. Also, when analytic decisions, such as anchor selection or tree splitting, are guided by the data, their uncertainty must be considered. Furthermore, statistical evidence of DIF does not necessarily imply that the difference is large enough to affect scores or decisions. Therefore, once DIF has been identified, we need further evidence to understand its source and choose an appropriate treatment.

\subsection{Extensions to ordinal and multidimensional measurement}

The binary 2PL model provides a simple setting in which to examine the identification problem, but many instruments have ordered or nominal response categories or measure several latent dimensions. For an ordinal item with categories $0,\ldots,M_j$, consider the cumulative-logit model
\begin{equation}\label{eq:ordinal}
 \logit\Prb(Y_{ij}\ge r\mid\theta_i,G_i=g)
 =a_j\theta_i-\kappa_{jr}+\delta_{jg,r},
 \qquad r=1,\ldots,M_j,
\end{equation}
where $\kappa_{jr}$ is a threshold parameter and  $\delta_{jg,r}$ is the group-specific deviation at threshold $r$, with $\delta_{j0,r}=0$. The
thresholds $\kappa_{jr}-\delta_{jg,r}$ must remain ordered within each group to ensure valid cumulative response probabilities. A common deviation
$\delta_{jg,r}=\delta_{jg}$ across all thresholds represents a shift in the cumulative response function without changing the relative spacing of the thresholds. Threshold-specific deviations allow the group contrast to differ across cumulative response probabilities. If the discrimination
parameter is also permitted to vary by group, the contrast may additionally depend on the latent trait. DIF in ordinal models must therefore be defined in relation to the parameters allowed to differ, whether these concern particular thresholds, the discrimination parameter or the response function as a whole. Regularised DIF estimation has been developed for generalised partial credit models with multiple candidate covariates \citep{schauberger2020}. When groups are latent, the identification problem also involves separating class-specific response parameters from differences in the latent trait distributions. \citet{wallinhuang2026ordinal} consider an ordinal latent-class formulation with class-specific intercept and slope differences, while \citet{wallinhuang2026asymmetric} examine joint latent impact and DIF under asymmetric item response functions. These extensions address different response-model restrictions without removing the need for common-scale identification.

Multidimensional measurement introduces a different complication. Let
$\boldsymbol\theta_i\in\mathbb R^Q$ represent the intended latent construct and $\mathbf a_j$ the corresponding item discrimination vector. A compensatory multidimensional model with group-specific item parameters can be written as
\begin{equation}\label{eq:multi}
 \logit\Prb(Y_{ij}=1\mid
 \boldsymbol\theta_i,G_i=g)
 =\mathbf a_j^\top\boldsymbol\theta_i+d_j
  +\delta_{jg}
  +\boldsymbol\gamma_{jg}^\top
   \boldsymbol\theta_i.
\end{equation}
The intercept deviation $\delta_{jg}$ represents uniform DIF conditional on the full latent trait vector, while $\boldsymbol\gamma_{jg}$ allows the
group contrast to vary with one or more latent dimensions. The interpretation of these deviations presupposes that the latent dimensions have been placed on a suitable common metric across groups.

Apparent DIF may also arise when an analysis conditions on an incomplete representation of the intended construct. To see this, suppose the item response function is invariant conditional on the full latent trait vector, but respondents are matched using a unidimensional variable $S_i$. The conditional response probability is then
\[
\begin{aligned}
 &\Prb(Y_{ij}=1\mid S_i=s,G_i=g)\\
 &\quad=
 \int
 \Prb(Y_{ij}=1\mid
       \boldsymbol\theta_i=\mathbf t)
 f(\mathbf t\mid S_i=s,G_i=g)
 \,d\mathbf t.
\end{aligned}
\]
Even when the item response function inside the integral is identical across groups, the resulting probabilities may differ if the conditional distributions of the latent dimensions differ. Matching on $S_i$ does not necessarily match respondents on the complete construct. \citet{ackermanma2024} examine these problems from a multidimensional IRT perspective, including the consequences of reducing a multidimensional latent space to a single matching dimension. The choice of matching variable and the specification of the latent construct are therefore
central to the interpretation of DIF. Introducing additional latent classes
cannot, by itself, resolve misspecification of the dimensions relevant to item responses.

\subsection{Effect sizes, instrument-level consequences and sensitivity analysis}

Statistical evidence of DIF does not, by itself, establish the magnitude or practical consequences of a difference in item functioning. Under Equation~\eqref{eq:general}, the conditional response-probability difference for item $j$ in group $g$, relative to the reference group, is
\begin{equation}\label{eq:pdiff}
 D_{jg}(\theta)
 =p_{jg}(\theta)-p_{j0}(\theta).
\end{equation}
Whereas the logit contrast $\Delta_{jg}(\theta)$ in Equation~\eqref{eq:difference} represents a difference in log odds, $D_{jg}(\theta)$ expresses the corresponding difference on the response-probability scale. The magnitude of the probability difference may vary across the latent trait even under uniform DIF, for which the log odds ratio is constant. Examining the item response functions therefore provides information that might not be apparent from a single estimated DIF coefficient.

Item-level effects can also be summarised over a specified range of the latent trait. Let $\mathcal T$ denote the trait interval of interest and let $w(\theta)$ be a non-negative weighting density satisfying $\int_{\mathcal T}w(\theta)\,d\theta=1$. Signed and absolute summaries are
\begin{equation}\label{eq:area}
 \begin{aligned}
 A_{jg}^{\mathrm{signed}}
 &=\int_{\mathcal T}
 D_{jg}(\theta)w(\theta)\,d\theta,\\
 A_{jg}^{\mathrm{abs}}
 &=\int_{\mathcal T}
 |D_{jg}(\theta)|w(\theta)\,d\theta.
 \end{aligned}
\end{equation}
The signed summary retains the direction of the group difference, whereas the absolute summary measures its magnitude without allowing positive and negative differences to cancel. Item-curve area measures have a longstanding
role in DIF research \citep{raju1988}. \citet{cichrova2026effects} examine the comparability of effect-size measures and classification thresholds across DIF procedures. Their results underscore why magnitude and classification should not be inferred from a significance test alone or from cutpoints borrowed uncritically across methods. Both the interval $\mathcal T$ and the weighting density affect the resulting summary. A common weighting distribution should therefore be specified for the
groups being compared, particularly when their latent trait distributions
differ. Weighting by the trait distribution of one group rather than another generally changes the population to which the summary is relevant.

Item-level DIF may also affect composite measurements. For an unweighted sum of binary item responses, the difference in the conditional expected composite between group $g$ and the reference group is
\begin{equation}\label{eq:dtf}
 \begin{aligned}
 D_{\mathrm{sum},g}(\theta)
 &=E\left(
   \sum_{j=1}^{J}Y_{ij}
   \mid\theta_i=\theta,G_i=g
   \right)\\
 &\quad-
   E\left(
   \sum_{j=1}^{J}Y_{ij}
   \mid\theta_i=\theta,G_i=0
   \right)\\
 &=\sum_{j=1}^{J}
   \{p_{jg}(\theta)-p_{j0}(\theta)\}.
 \end{aligned}
\end{equation}
This identity does not require conditional independence of the item responses. For a weighted composite, each item-level difference is multiplied by its corresponding weight.

The conditional composite effect depends on both the magnitude and direction of the item-level effects. Differences that favour opposite groups may offset one another, whereas smaller effects in the same direction may accumulate. Cancellation in the expected composite does
not establish that the individual items are acceptable for the
intended measurement purpose. Differences in particular content
domains or item response patterns could remain important, even when
their contributions to the composite mean are small. The distinction between item-level and composite-level effects underlies the established
concept of differential test functioning and applies more generally to questionnaire totals, scale scores and other composite measurements.

The conditional composite effect in Equation~\eqref{eq:dtf} should be distinguished from differences in observed group means. It compares expected composite scores at the same latent trait value and therefore describes the contribution of DIF under the specified model. Observed group mean differences also reflect differences in the groups' latent trait distributions. Consequently, groups may have different average composite scores even when every item is invariant, and the absence of an observed mean difference does not establish the absence of DIF.

Sensitivity analysis is particularly important when the invariant measurement structure is uncertain. Estimated DIF effects may change when different anchor sets, matching variables or identifying restrictions are used. For latent-group models, additional sensitivity may arise from the assumed number of classes, latent trait distributions and regularisation parameters. Comparing results across plausible specifications can reveal whether the conclusions depend strongly on particular modelling
decisions. Such comparisons cannot establish which specification represents
the underlying response process, but they help determine which conclusions are supported under a range of defensible assumptions. Particular attention is warranted when the identity of the items displaying DIF changes across alternative common metrics. 

\subsection{Evidence for DIF sources and implications for item treatment}

The detection of DIF establishes a conditional group difference, but it does not indicate why the difference occurs or how the item should be treated. Some DIF can be traced to wording and translation. Content review and cognitive interviews can help identify these problems and guide item revision. Other cases may reflect differential familiarity with the item context or a secondary skill. Explanatory IRT or mediated MIMIC models can examine proposed item features, while a multidimensional model could assess the role of the secondary skill. The response model specification itself also deserves attention. For example, \citet{huangboltlyu2024} showed that fitting a symmetric IRT model (e.g., 2PL) to an item with an asymmetric response function can yield artificial DIF across groups. In their TIMSS application, allowing item-specific asymmetry reduced some of the detected DIF. 

With the development of computer-based assessments, additional information through responses times and actions sequences become available as tools to diagnose, interpret, and reduce DIF. These data can reveal group differences in how an item is approached that are not apparent from the final response \citep{ercikan2020}. \citet{li2025processdif}, for example, used process features to examine differences in response efficiency and exploration for items displaying gender DIF. \citet{molenaarfeskens2026} went further by specifying a latent response-process factor and testing whether group differences on that factor could account for violations of measurement invariance. \citet{chen2026processdif} constructed a surrogate for a nuisance trait and proposed a scoring rule intended to reduce DIF attributable to that trait. These studies use process data for different purposes: interpreting a detected difference, modelling a possible response process, and modifying scoring.

Let $Z_{ij}$ denote an observed process indicator for respondent $i$ on item $j$. In parallel with Equation~\eqref{eq:pdiff}, the group contrast conditional on both proficiency and the process indicator is
\begin{equation}
D_{jg}(\theta,z)
=
\Pr(Y_{ij}=1\mid\theta_i=\theta,G_i=g,Z_{ij}=z)
-
\Pr(Y_{ij}=1\mid\theta_i=\theta,G_i=0,Z_{ij}=z).
\label{eq:process_conditional_dif}
\end{equation}
This contrast can show whether a group difference remains among respondents with similar values of the recorded indicator, provided those values are represented in both groups. However, a smaller group contrast after conditioning on $Z_{ij}$ does not establish that the recorded behaviour caused the DIF. Moreover, $D_{jg}(\theta,z)$ and $D_{jg}(\theta)$ condition on different information, so their difference cannot be interpreted as the amount of DIF explained by $Z_{ij}$. A short response time, for example, may indicate fluency, guessing, or disengagement. Interpreting the indicator therefore requires evidence about the response process and its relationship to the intended construct \citep{goldhammerzehner2017}.

Finally, the decision about how to handle a DIF item cannot be made from the DIF statistic alone. Removing an item may eliminate a suspected source of bias but also reduce coverage of the content. Keeping the item with group-specific parameters preserves information for within-group scoring, although it changes how the item contributes to cross-group comparisons. Studies comparing these and other treatments show that the choice can affect item calibration, person scores, and estimates of group impact \citep{cho2016afterdif, liurogers2022}. Their consequences should be compared before deciding how the item will contribute to scoring.

\section{Discussion}\label{sec:discussion}

DIF methods differ in how they match respondents, establish a common measurement scale and represent population heterogeneity. Their connections are apparent once the scale is treated as part of the analysis rather than assumed to be given. The responses used to estimate a construct may themselves exhibit DIF, so separating differences in the construct from differences in item functioning requires identifying assumptions.

The distinction between observed and latent groups does not imply a progression from simpler to more appropriate methods. Observed group comparisons address substantively defined populations; recursive partitioning can discover relevant combinations of recorded characteristics; mixture models represent heterogeneity that is not captured by those characteristics. Likewise, selecting anchors, imposing approximate invariance and using robust scaling are different responses to uncertainty about the common metric. A method is appropriate only insofar as its comparison and identifying restrictions correspond to the intended measurement question.

One unresolved problem concerns the behaviour of anchor-free procedures when DIF is widespread or predominantly in one direction. Such patterns may violate the assumptions used to establish a common metric without prespecified invariant items. The location-shift equivalence explains the underlying difficulty: a component of the item differences proportional to their discriminations can be attributed either to item functioning or to a group-specific shift in the latent trait. Sparse identification, approximate invariance and robust scaling resolve this ambiguity through different assumptions about the pattern of item effects. When plausible identifying restrictions lead to different latent mean contrasts or DIF classifications, that sensitivity should be reported rather than resolved solely by selecting the model with the most favourable fit statistic.

Latent-group analysis adds uncertainty about how the population is represented. Mixture components may accommodate a non-normal trait distribution, local dependence, omitted dimensions or class-specific item response functions. The equivalence between an invariant mixture and a single-class model with the corresponding mixture trait distribution illustrates why the presence of classes does not establish DIF. Evaluation of joint methods should consider the common metric, item effects, trait distributions and membership probabilities together, including settings with weak separation, class imbalance and misspecified response models. Numerical convergence alone gives limited reassurance if different
initialisations produce different item classifications or group interpretations.

Statistical inference after data-dependent identification remains a separate issue. Selecting anchors, tuning penalties or choosing a class structure can change the sampling distribution of later estimates. Refitting after selection may reduce shrinkage bias without accounting for that selection. The framework of \citet{chen2023} provides one treatment of identification and inference with observed groups and unknown anchors; simultaneous selection of classes and invariant items presents further challenges. More general approaches, including those that can handle unobserved groups, remain an open problem.

In applications, a detected item difference must be considered alongside its magnitude, the trait range over which it occurs and its consequences for measurement or intended decisions. DIF does not determine the reason an item functions differently or whether it should be modified or excluded. Content, response processes and the instrument's purpose remain relevant;
for latent classes, external information may be needed even to interpret which respondents the groups represent. The common methodological task is to state what information makes the comparison possible, what uncertainty remains and which conclusions are supported under plausible alternative measurement structures.

\newpage

\bibliographystyle{apalike}
\bibliography{references}

\end{document}